# Manipulating excitons in MoSe$_2$ with 2D ferroelectric gating

Xiaoyu Mao[1,2,#], Jun Fu[1,2,#], Ming Gong[3,4,*] and Hualing Zeng[1,2,*]

1. International Center for Quantum Design of Functional Materials (ICQD), Hefei National Laboratory for Physical Science at the Microscale, and Synergetic Innovation Center of Quantum Information and Quantum Physics, University of Science and Technology of China, Hefei, Anhui 230026, China

2. Key Laboratory of Strongly-Coupled Quantum Matter Physics, Chinese Academy of Sciences, Department of Physics, University of Science and Technology of China, Hefei, Anhui 230026, China

3. CAS Key Laboratory of Quantum Information, University of Science and Technology of China, Hefei, 230026, People's Republic of China

4. Synergetic Innovation Center of Quantum Information and Quantum Physics, University of Science and Technology of China, Hefei, Anhui 230026, China

\# Contribute equally to this work

\* Corresponding author

## Introduction

Ever since 2004 Novoselov and Geim et al.[1] prepared graphene through mechanical exfoliation, more and more atomically thin Van der Walls materials have approached into the scene view of researchers. For the small size and potential application in logical circuits, semiconductor materials with a single atomic layer thickness have attracted much attention. 2010 KF Mak et al.[2] found that for the disappearance of interlayer coupling, MoS$_2$ exhibit a transition from indirect to direct bandgap in the limit of monolayer thickness. This indirect-to-direct bandgap transition leading to a huge enhancement in photoluminescence (PL) quantum yield (QY). Since then it has been discovered that monolayer two-dimensional transition metal chalcogenides (TMDCs) have similar characteristics[3,4]. On the other hand, contrast to bulk, the excitons in monolayer TMDCs are confined in the 2D plane with reduced screening due to the change of dielectric environment, resulting in a greater binding energy[5,6].

For the band gap within the visible range and its natural two-dimensional properties, monolayer TMDCs provide a platform for the study of band engineering and exciton effects of two-dimensional materials. $MoSe_2$, as a member of TMDCs family, has a sharper PL spectrum at low temperature, so it can show a clearer picture for studies of exciton effects and excitons behavior in external fields. 2013 Jason et al.[7] obtained monolayer $MoSe_2$ on $SiO_2$ substrate using mechanical exfoliation method. By fabricating field-effect transistors (FET) device, they respectively achieved *n* doping and *p* doing in monolayer $MoSe_2$. Excitons and both positive and negative trions was observed, and the excitons and trions PL peak intensity can be controled by changing the back gate voltage. However, this FET structure can keep the doping level only by maintaining the back-gate voltage. Replacing $SiO_2$ with ferroelectric materials seems to be a neutral idea to optimize. 2018 Bo Wen et al.[8] stacked Monolayer $MoSe_2$ on the surface of $LiNbO_3$. They observed opposite doping types of $MoSe_2$ at different domain in $LiNbO_3$. Whereas, there is depolarization field in traditional ferroelectric crystal such as $LiNbO_3$ and $BaTiO_3$ et al. The depolarization field will be strong enough to cancel the ferroelectric polarization when the thickness is reduced to a certain value, which is called critical size effect[9-11]. Two-dimensional Van der Walls materials have natural two-dimension properties and saturated surface chemical states, so they offer an opportunity to find ferroelectricity in two dimensional limit[12]. There is a natural spatial symmetry break of spatial inversion in $CuInP_2S_6$, and the bulk $CuInP_2S_6$ has been proved to be room temperature ferroelectric material. In 2016, further study proofed that 4nm $CuInP_2S_6$ can maintain its ferroelectric properties at room temperature[13].

On the other hand, the layer structure of these Van der Walls materials provide us a chance to stack different kinds of materials together to obtain heterostructures like Lego toys[14], which makes it is possible for us to use two-dimensional materials as gate to modulate the PL characteristic of $MoSe_2$. Dry transfer method can avoid the introduction of impurities and defects in organic solvents, so we fabricated $MoSe_2$/$CuInP_2S_6$ heterostructure devices using an all dry transfer method[15]. By polarizing $CuInP_2S_2$ with different voltage, we can manipulate the excitons behavior in $MoSe_2$ with two-dimensional material.

# Result

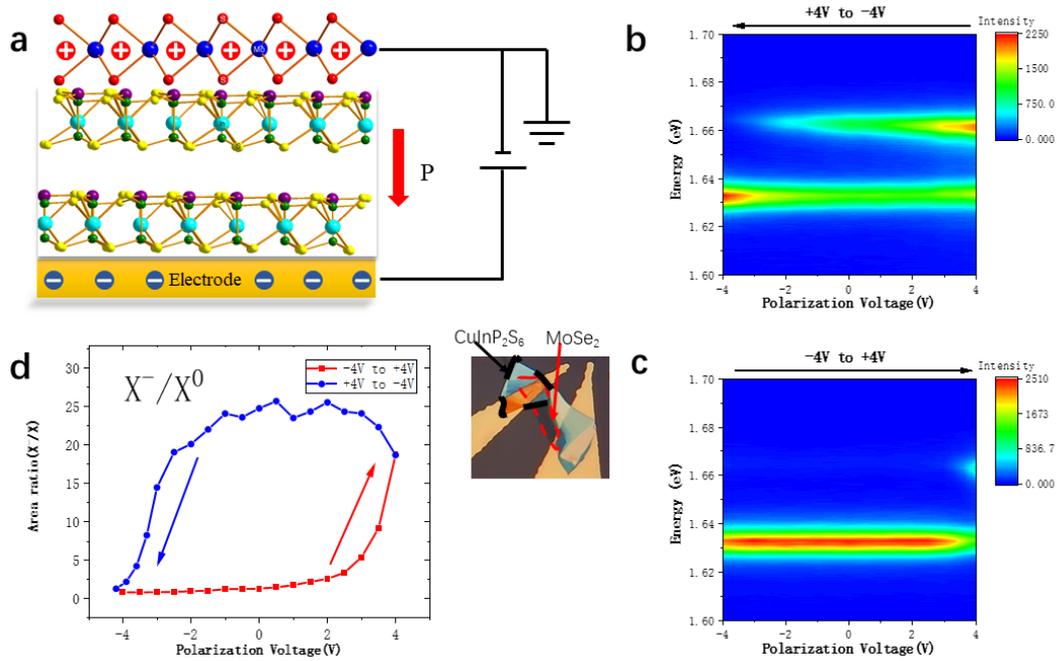

**Figure 1.** MoSe$_2$/CuInP$_2$S$_6$ device structure and non-volatile modulation. a) Structure schematic of ferroelectric plane-parallel capacitor. The larger arrow on the right shows the direction of electric polarization at negative voltageInsert: the optical photograph of the device, the substrate is Si wafer with 90 nm fused SiO$_2$ on top. The few-layers CuInP$_2$S$_6$ and monolayer MoSe$_2$ are indicated by black and red dashed frames, respectively. b) PL mapping of MoSe$_2$ as polarization voltage change from +4V to -4V and c) -4V to +4V. d) PL intensity ratio ($X^-/X^0$) hysteresis induced by polarization voltage variation.

The device structure schematic is depicted in Figure 1a. Few-layer CuInP$_2$S$_6$ and monolayer MoSe$_2$ were prepared on polydimethylsiloxane (PDMS) by mechanical exfoliation method. Using an all-dry transfer method, few-layer CuInP$_2$S$_6$ was transferred on the surface of Au electrode. Monolayer MoSe$_2$ was stacked on CuInP$_2$S$_6$ in the same way. By contacting MoSe$_2$ with another electrode during transfer, we can polarize CuInP$_2$S$_6$ by applying polarization voltage between these two electrodes. Insert is the optical photograph of device. In order to prevent light doping and other kinds of excitons generation due to high-power excitation, the laser power of 10μW was used to measure the PL spectrum. For preventing the impact of grounding on the measurement, we disconnected the low potential end after polarization, and we used a pulse voltage to polarize CuInP$_2$S$_6$ better. All the PL measurement were performed at 7K if not specifically mentioned. With little impacts of thermal excitation at 7K, electron-phonon interaction becomes extremely weak, which makes the PL peak is quiet sharp. Typical PL spectrum without electric control is displayed in Figure S1 in Supporting Information. There is almost no wide defect peak in the PL spectrum of MoSe$_2$, which indicates the high quality of the MoSe$_2$ single crystal. Figure 1b and 1c shows PL color maps of the device as polarization voltage changing from +4V to -4V and back. Two very sharp PL peak observed at 1.66eV and 1.63eV correspond to neutral excitons and charged trions in monolayer MoSe$_2$. The energy difference about 30meV between the two peaks matches trion binding energy. Excitons can form positive or negative trions by binding an additional electron ($X^-$) or hole ($X^+$). After +4V polarization, we can observe both excitons and trions PL peak clearly in spectrum. While after -4V polarization, CuInP$_2$S$_6$'s surface provides plenty of carriers, resulting in heavily doping in MoSe$_2$. The high doping level in MoSe$_2$ leads to electron screening effect, hence the excitons emission is strongly suppressed. We mainly observed trions PL peak. During the variation of voltage, we can observe that when the voltage goes from negative to positive and back to negative, all the PL spectrum of MoSe$_2$ have hysteresis. We integrated the area of excitons and trions PL peak to obtain the peak intensity X$^0$ and X$^-$. Figure 1d shows the ratio of X$^-$/X$^0$ as a function of polarization voltage. We find that the dependence between this ratio and voltage have the similar shape of the hysteresis line, which indicates that CuInP$_2$S$_6$ can be effectively polarized in this configuration while achieving non-volatile modulation of monolayer MoSe$_2$. We also performed room temperature PL spectrum versus polarization at 297K (Figure S2 in Supporting Information). Since the electron-phonon interaction and thermal excitation of bound electrons in trions, we can't divide excitons and trions peak. But by extracting the peak intensity dependence on polarization voltage, we can also obtain a hysteresis shape, indicating the room temperature ferroelectricity of CuInP$_2$S$_6$ flakes.

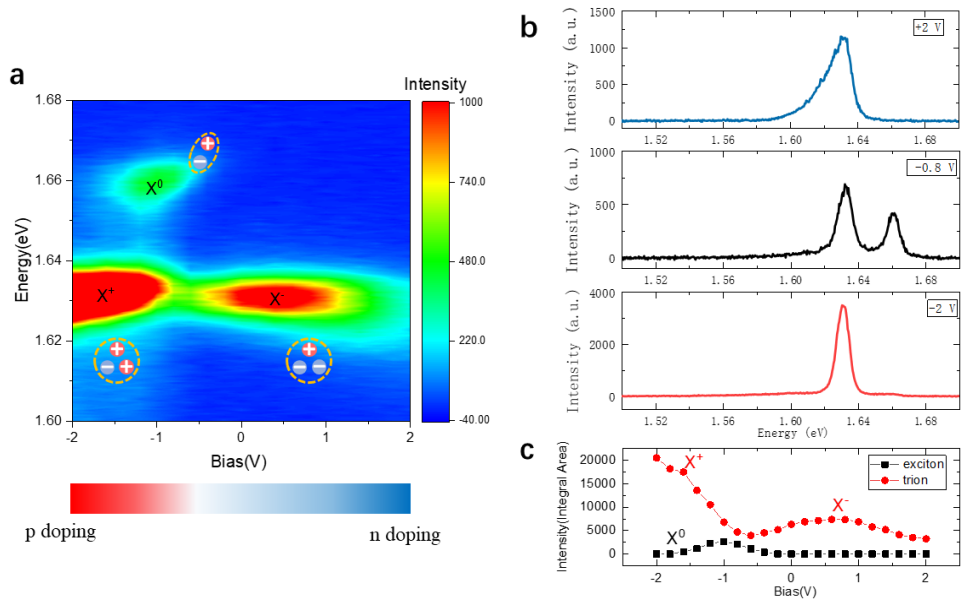

Figure 2. MoSe$_2$ PL dependence on direct current (DC) mode voltage. a) MoSe$_2$ PL mapping as the DC voltage change from -2V to +2V. The excitons species are represented by hydrogen and hydrogen ion like models shown in picture. b) Characteristic PL peak at +2V, -0.8V and -2V represent for different doping types and doping level. c) Excitons (X$^0$) and trions peak intensity as a function of bias.

To better distinguish the excitons species in MoSe$_2$, a direct current (DC) mode voltage test was performed at 7K. Figure 2a shows the PL mapping during the DC voltage changing from -2V to 2V. Figure 2b shows the representative PL spectrum at +2V, -0.8V and -2V for different doping types. For the similar structure with capacitor and very small thickness of few-layers CuInP$_2$S$_6$, DC voltage can achieve efficiently carrier doping for MoSe$_2$. In the voltage change process from +2V to -0.8V, MoSe$_2$ is mainly doped with electrons. The excitons PL peak intensity is almost zero. We can mainly observe trions peak. As shown in Figure 2a, under positive voltage modulation (electron doping), trions peak broadens to the low energy direction. Unlike excitons, trions can radiatively decay with non-zero momentum by kicking out an electron. For energy and momentum conservation, the emitted photon energy for trions is given by $\hbar\omega_T = E_{trion}^0 - \frac{M_X}{M_e}E_{KE}$, where $E_{trion}^0$ is the zero-momentum trions enegy, $M_X$ is exciton mass, $M_e$ is the effective electron mass, $E_{KE}$ is electrion kinetic energy (The schematic is shown in Figure S3 in supporting information). This electron recoil effect[16] results in the long tailed trions peak. As the positive voltage going up, the broadening of the trions peak is more obvious. At the same time, the overall PL intensity decrease due to the electron screening of excitons at higher electron concentration. At near -0.8V, the carriers doping is relatively weak, and the trions intensity is greatly reduced. The excitons peak can be clearly observed because there is few free carriers' suppression. But when the voltage goes towards -2V direction, due to the introduction of holes, the PL intensity of neutral excitons is suppressed, and the positively charged trions are significantly enhanced. At the same time, due to the recoil effect like the electron, we can observe the X$^+$ peak broadens to low-energy direction. We extract the excitons (black) and trions (red) peak intensity as a function of DC voltage in Figure 2c. The X$^0$ and trions intensity show competitive relation, for the carriers suppression and screening on excitons. In some other devices, such as shown in Figure 3a, the excitons peak won't be suppressed completely at negative voltage, which may be caused by the incomplete reversal polarization state of $CuInP_2S_6$.

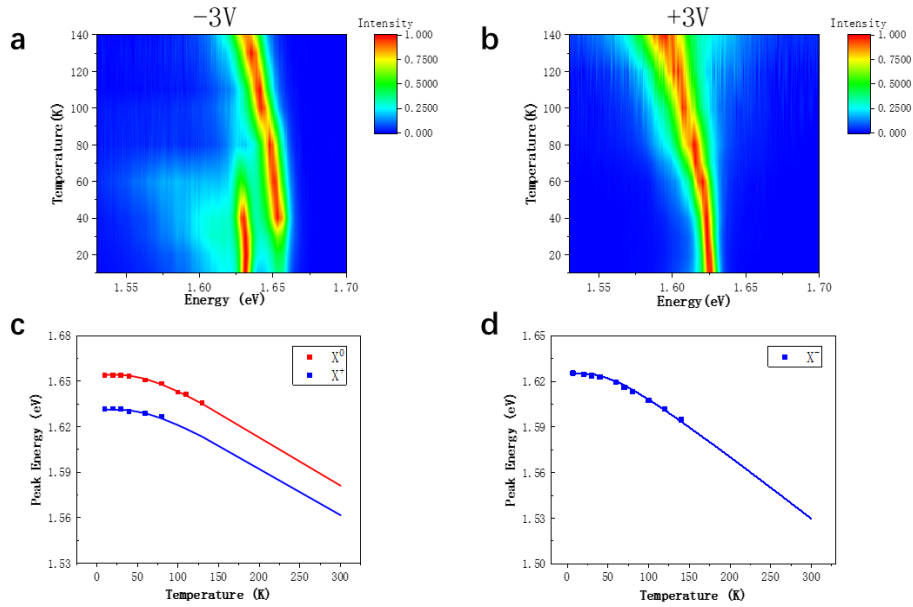

Figure 3. Temperature dependence of $MoSe_2$ PL spectrum. a) Normalized PL against temperature at -3V and b) +3V. c) Neutral excitons (X$^0$) (red) and positive trion (X$^+$) (blue) peak position against temperature at -3V d) Negative trion (X$^-$) peak position against temperature at +3V.

Figure 3a shows the $MoSe_2$ PL mapping as a function of temperature at -3V (DC). We can see three peaks in the PL spectrum at low temperature. The two sharp peaks are of excitons and positive trions (X$^+$), and the broad low energy peak at around 1.6eV may be caused by defect. The defect peak disappears above 80K, maybe because the electrons escape from the defect energy level. And the trions peak intensity drops apparently at around 60K, which can be explained by that higher temperature provide thermal energy for holes to escape from their bound trion state. We fit the PL peak with multiple Lorentz function and extract the phonon energies of $X^0$ and $X^+$, which as shown in Fig 3c. We don't present the $X^+$ data above 80K because it becomes negligible. These two peaks' position fit well (solid line in Figure 3c) by the O'Donnell's equation[17], an equation widely used to describe the temperature dependence of semiconductor bandgap:

$$E_g(T) = E_g(0) - S\hbar\omega \left[\coth\left(\frac{\hbar\omega}{2kT} - 1\right)\right]$$

Where $E_g(0)$ is the ground state transition energy at 0K, $S$ is a dimensionless constant describing the strength of the electron-phonon coupling, and $\hbar\omega$ represents the average acoustic phonon energy involved in electron-phonon interaction. From the fit, we extract for X($X^+$) $E_g(0) = 1.654\ (1.631)\ \text{eV}$, $S = 1.89\ (2.1)$ and $\hbar\omega = 16.8\ (20)\ \text{meV}$. Figure 3b shows the same PL mapping at +3V (DC voltage). For heavily electron doping in MoSe2, neutral excitons' PL peak is strongly suppressed, only $X^-$ peak is observed. We can also extract the peak position fitting with Lorentz function. The temperature dependence of peak position is depicted in Figure 3d, which can be fitted by O'Donnell's equation (the solid line in Figure 3d). For $X^-$ $E_g(0) = 1.625\ \text{eV}$, $S = 2.02$, $\hbar\omega = 11.0\ \text{meV}$. The slight ground state transition energy difference between $X^+$ and $X^-$ may be caused by the different binding energy of excitons binding an electron or a hole.

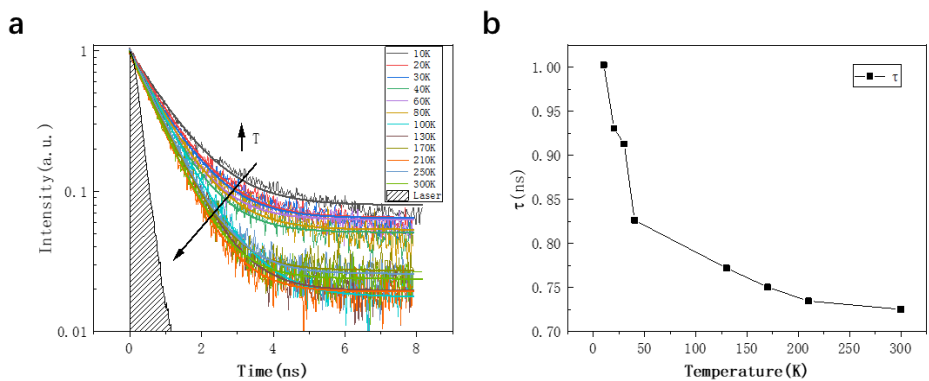

Figure 4. Temperature dependence time-resolved photoluminescence spectrum. a) Normalized time-resolved photoluminescence against temperature. b) Excitons radiative decay lifetime against temperature.

We also investigated the radiative properties by time-resolved photoluminescence (TRPL) (detailed setup and methods are in Experimental Section). We first tested the TRPL spectrum versus voltage at 10K (Figure S4 Supporting information), in which we can extract that the excitons radiative decay lifetime are almost the same at the voltage range from -4V to +4V (DC). Then we focused on the TRPL spectrum versus temperature at -3V. Because we can observe both neutral excitons and charged trions peak at -3V, from which we can investigate dynamics of both. Figure 4a displays the MoSe$_2$/CuInP$_2$S$_6$ heterostructure's TRPL spectrum from 10K to 300K at -3V (DC). We can fit the curve with single exponential function, which means there is only one radiation channel. What surprised us is that the lifetime of excitons is all several hundred of picoseconds at the full temperature range, which is apparently much longer than previous examination of TMDCs. On the other hand, the excitosn lifetime is obviously longer than the resolution of our setup (the shadow area in Figure 4a). It is all known that excitons luminescence must satisfy conservation of energy and momentum, so only the excitons in the light cone can radiatively recombine. In a simple approach, the exciton intrinsic radiative decay time is $\tau_0 = \frac{1}{2\Gamma}$, where $\Gamma$ is the radiative decay rate. $\tau_0$ for monolayer TMDCs is 3-5 ps at 10K typically[18,19]. We attribute this anomaly to two factors. First, the excitation light energy 2.25eV is much bigger than the bandgap of MoSe$_2$, which result in a long-time relaxation. The relaxation time masks the intrinsic excitions radiative lifetime. Second, the MoSe$_2$/CuInP$_2$S$_6$ interlayer coupling provides a nonradiative channel for excitons[20]. The nonradiative recombination process comes from the exciton-exciton annihilation process (Auger type), and may dominate the decay time. As shown in Figure 4b, exciton lifetime decrease as temperature goes up, which is in agreement with that the excitons' mobility increases with temperature.

## Experimental Section

*Sample preparations and device fabrications*: The bulk single crystal CuInP$_2$S$_6$ in this study was purchased from HQ Graphene, Inc. The bulk single crystal MoSe$_2$ was synthesized by chemical vapor transportation method. The ultrathin CuInP$_2$S$_6$ and monolayer MoSe$_2$ were obtained by mechanical exfoliation on PDMS (Gel-Pak, WF-60-X4). The heterostructures were made by an all dry transfer method as described in ref [15]. Ultrathin two-dimensional materials on PDMS were transferred on prepared electrodes on SiO$_2$/Si wafer under a microscope with the help of a home-made XYZ mechanism. The specific transfer process is shown in Figure S5 in Supporting Information. The electrodes were made by standard photoetching method and deposition 5 nm titanium and 25 nm gold via e-beam evaporator.

*Optical measurements.* Photoluminescence measurements were performed via Princeton Instruments spectrometer (IsoPlane® SCT 320) with 532nm laser. The laser was focused by a 100X objective to a spot (~2μm) on the device, the laser intensity was kept below 10μW. For the TRPL measurements, the supercontinuum light was generated from ultrafast laser source (MaiTai HP, <100fs) using a photonic crystal fiber, then passed through a tunable lowpass filter to obtain 530nm excitation radiation, and finally focused on the sample. The reflected signal was collected and analyzed by a Time-Correlated Single Photon Counting (PicoQuant TimeHarp 260), which has 25ps resolution. For both PL and TRPL measurements, the device of MoSe$_2$/CuInP$_2$S$_6$ heterostructure was placed in a cryogenic chamber (Montana Instruments) capable of reaching temperatures down to 4 K

# References

1　Novoselov, K. S. *et al.* Electric field effect in atomically thin carbon films. *science* **306**, 666-669 (2004).

2　Mak, K. F., Lee, C., Hone, J., Shan, J. & Heinz, T. F. Atomically thin MoS(2): a new direct-gap semiconductor. *Phys Rev Lett* **105**, 136805, doi:10.1103/PhysRevLett.105.136805 (2010).

3　Tongay, S. *et al.* Thermally driven crossover from indirect toward direct bandgap in 2D semiconductors: MoSe2 versus MoS2. *Nano Lett* **12**, 5576-5580, doi:10.1021/nl302584w (2012).

4　Zhao, W. *et al.* Evolution of Electronic Structure in Atomically Thin Sheets of WS2 and WSe2. *ACS Nano* **7**, 791-797, doi:10.1021/nn305275h (2013).

5　Chernikov, A. *et al.* Exciton binding energy and nonhydrogenic Rydberg series in monolayer WS(2). *Phys Rev Lett* **113**, 076802, doi:10.1103/PhysRevLett.113.076802 (2014).

6　Raja, A. *et al.* Coulomb engineering of the bandgap and excitons in two-dimensional materials. *Nature communications* **8**, 15251 (2017).

7　Ross, J. S. *et al.* Electrical control of neutral and charged excitons in a monolayer semiconductor. *Nat Commun* **4**, 1474, doi:10.1038/ncomms2498 (2013).

8　Wen, B. *et al.* Ferroelectric-Driven Exciton and Trion Modulation in Monolayer Molybdenum and Tungsten Diselenides. *ACS Nano* **13**, 5335-5343, doi:10.1021/acsnano.8b09800 (2019).

9　Trieloff, M. *et al.* Structure and thermal history of the H-chondrite parent asteroid revealed by thermochronometry. *Nature* **422**, 502-506, doi:10.1038/nature01499 (2003).

10　Fong, D. D. *et al.* Ferroelectricity in Ultrathin Perovskite Films. *Science* **304**, 1650-1653, doi:10.1126/science.1098252 (2004).

11　Gao, P. *et al.* Possible absence of critical thickness and size effect in ultrathin perovskite ferroelectric films. *Nat Commun* **8**, 15549, doi:10.1038/ncomms15549 (2017).

12　Ding, W. *et al.* Prediction of intrinsic two-dimensional ferroelectrics in In2Se3 and other III2-VI3 van der Waals materials. *Nat Commun* **8**, 14956, doi:10.1038/ncomms14956 (2017).

13　Liu, F. *et al.* Room-temperature ferroelectricity in CuInP 2 S 6 ultrathin flakes. *Nature communications* **7**, 12357 (2016).

14　Geim, A. K. & Grigorieva, I. V. Van der Waals heterostructures. *Nature* **499**, 419-425, doi:10.1038/nature12385 (2013).

15　Castellanos-Gomez, A. *et al.* Deterministic transfer of two-dimensional materials by all-dry viscoelastic stamping. *2D Materials* **1**, 011002 (2014).

16　Christopher, J. W., Goldberg, B. B. & Swan, A. K. Long tailed trions in monolayer MoS2: Temperature dependent asymmetry and resulting red-shift of trion photoluminescence spectra. *Sci Rep* **7**, 14062, doi:10.1038/s41598-017-14378-w (2017).

17　O'Donnell, K. P. & Chen, X. Temperature dependence of semiconductor band gaps. *Applied Physics Letters* **58**, 2924-2926, doi:10.1063/1.104723 (1991).

18　Korn, T., Heydrich, S., Hirmer, M., Schmutzler, J. & Schüller, C. Low-temperature photocarrier dynamics in monolayer MoS2. *Applied Physics Letters* **99**,

        doi:10.1063/1.3636402 (2011).
19      Wang, G. *et al.* Valley dynamics probed through charged and neutral exciton emission in monolayerWSe2. *Physical Review B* **90**, doi:10.1103/PhysRevB.90.075413 (2014).
20      Robert, C. *et al.* Exciton radiative lifetime in transition metal dichalcogenide monolayers. *Physical Review B* **93**, doi:10.1103/PhysRevB.93.205423 (2016).